\documentclass[twocolumn,letter]{jpsj3}
\usepackage{graphicx}
\usepackage{dcolumn}
\usepackage{bm}
\usepackage{ulem}
\usepackage{braket}
\usepackage{amsmath}
\usepackage{color}

\title{Multiple skyrmion crystal phases by itinerant frustration in centrosymmetric tetragonal magnets}

\author{Satoru Hayami}
\inst{
Department of Applied Physics, The University of Tokyo, Tokyo 113-8656, Japan
 }

\abst{
A skyrmion crystal (SkX) expressed as a multiple number of spiral modulations manifests itself not only in its peculiar magnetic texture but also in nontrivial transport properties originating from an emergent magnetic field. 
We here report our numerical results for multiple SkXs in a centrosymmetric tetragonal crystal system. 
By performing simulated annealing for an effective spin model for itinerant magnets, we find that three types of the SkXs with the skyrmion numbers of one and two, which are characterized by different superpositions of helices, are stabilized in the ground state, and are transformed by an external magnetic field. 
The essence of the emergent multiple SkXs is lied in itinerant frustration where exchange interactions are competed in momentum space due to the nature of itinerant electrons. 
}

\begin{document}
\maketitle

A multiple-$Q$ magnetic ordering, which is constructed from a superposition of helices, has been extensively studied in recent years, since its spin texture is related to a topological spin texture that causes the topological Hall effect and multiferroic phenomena~\cite{nagaosa2013topological,batista2016frustration,fert2017magnetic,Tokura_doi:10.1021/acs.chemrev.0c00297}.
Depending on a way of superposing helices, various topological spin textures, such as a magnetic skyrmion, vortex, meron, and hedgehog, are realized~\cite{Bogdanov89,Bogdanov94,rossler2006spontaneous,Binz_PhysRevB.74.214408,Yi_PhysRevB.80.054416,hayami2021topological}. 
Notably, such multiple-$Q$ orderings have been ubiquitously found in both noncentrosymmetric~\cite{Muhlbauer_2009skyrmion,yu2010real,yu2011near,seki2012observation,tanigaki2015real,nayak2017discovery,kakihana2018giant,tabata2019magnetic,kaneko2019unique,fujishiro2019topological,hayami2021field} and centrosymmetric~\cite{Saha_PhysRevB.60.12162,takagi2018multiple,kurumaji2019skyrmion,hirschberger2019skyrmion,Ishiwata_PhysRevB.101.134406,khanh2020nanometric,Yasui2020imaging,seo2021spin,Hirschberger_10.1088/1367-2630/abdef9} magnets composed of $p$, $d$, and $f$ electrons, although their mechanisms are qualitatively different: The former relies on the Dzyaloshinskii-Moriya (DM) interaction~\cite{dzyaloshinsky1958thermodynamic,moriya1960anisotropic} and the latter is accounted for by frustrated exchange interactions~\cite{Okubo_PhysRevLett.108.017206,leonov2015multiply,Lin_PhysRevB.93.064430,Hayami_PhysRevB.93.184413,Hayami_PhysRevB.94.174420,mitsumoto2021replica,Hayami_PhysRevB.103.224418} and/or effective magnetic interactions mediated by itinerant electrons~\cite{Akagi_PhysRevLett.108.096401,Hayami_PhysRevB.90.060402,Ozawa_doi:10.7566/JPSJ.85.103703,Ozawa_PhysRevLett.118.147205,Hayami_PhysRevB.95.224424,Wang_PhysRevLett.124.207201,Eto_PhysRevB.104.104425}. 
Their different mechanisms lead to a difference of magnetic modulation periods in the multiple-$Q$ orderings; the latter mechanism tends to favor the short-period skyrmion crystal (SkX) compared to the former one, which might be promising for high-efficient spintronic devices consisting of high-density topological objects~\cite{Zhang_2020}. 

\begin{figure}[t!]
\begin{center}
\includegraphics[width=1.0\hsize]{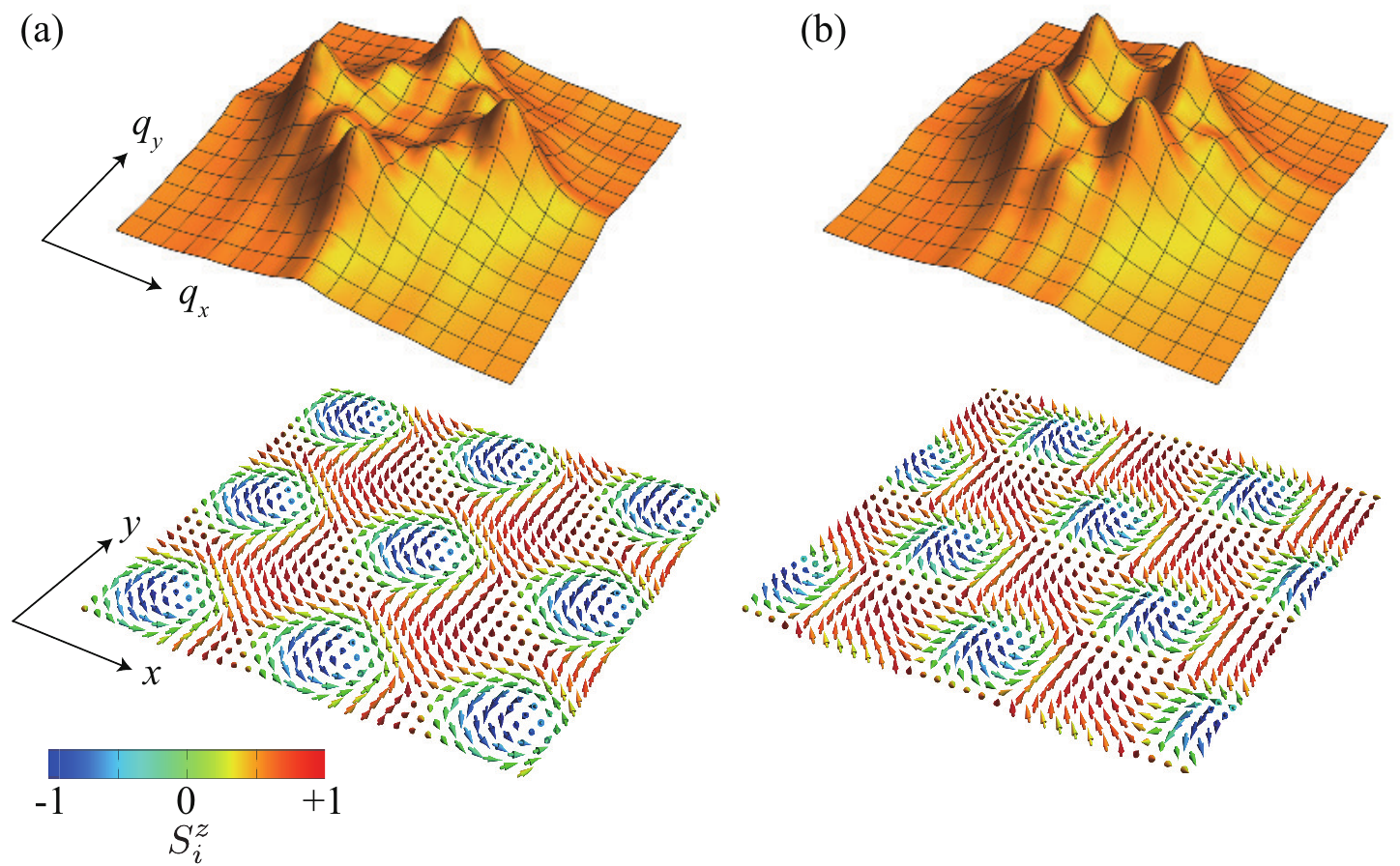} 
\caption{
\label{fig:ponti}
(Upper panel) Schematics of the interactions in momentum space with maxima at (a) $\bm{Q}_1=(Q,0)$ and $\bm{Q}_2=(0,Q)$ and (b) $\bm{Q}_3=(Q',Q')$ and $\bm{Q}_4=(-Q',Q')$
(Lower panel) SkXs constructed from the superposition of helices with (a) $\bm{Q}_1$ and $\bm{Q}_2$ or (b) $\bm{Q}_3$ and $\bm{Q}_4$. 
}
\end{center}
\end{figure}

For the latter mechanisms without relying on the DM interaction, magnetic spiral periods are determined by exchange interactions in momentum space.  
For instance, the periods in frustrated Mott insulators are set by the Fourier transform of the exchange interactions $J_{\bm{q}}=\sum_{ij}J_{ij}e^{-i\bm{q}\cdot (\bm{r}_i-\bm{r}_j)}$, where $J_{ij}$ is the exchange coupling between spins at sites $i$ and $j$, $\bm{q}$ is the wave vector in momentum space, and $\bm{r}_i$ is the position vector for site $i$. 
Similarly, the periods in itinerant magnets consisting of itinerant electrons and localized spins are determined by the Ruderman-Kittel-Kasuya-Yosida (RKKY) interaction $J_{\rm K}^2 \chi_{\bm{q}}$~\cite{Ruderman,Kasuya,Yosida1957}, where $J_{\rm K}$ is the exchange coupling between itinerant electron spins and localized spins and $\chi_{\bm{q}}$ is the bare susceptibility of itinerant electrons with the wave vector $\bm{q}$. 
The important common feature is that there are multiple choices of the ordering vectors, which are connected by the rotational symmetry of the lattice structure, since the helix with one of the ordering vectors is energetically degenerate with the symmetry-related helices, as shown in the upper panel of Fig.~\ref{fig:ponti}. 
In this situation, additional interactions and fluctuations stabilize the multiple-$Q$ states instead of the single-$Q$ helical state~\cite{batista2016frustration,hayami2021topological}. 
We show the example of the square SkXs consisting of the double-$Q$ spiral modulations along the $\langle 100\rangle$ and $\langle 110\rangle$ directions in the lower panel of Figs.~\ref{fig:ponti}(a) and \ref{fig:ponti}(b), respectively.
In particular, in itinerant magnets, the relatively large strength of multiple-spin interactions beyond the RKKY level brings about the instability toward the multiple-$Q$ states, which is in contrast to the insulating magnets with the short-ranged competing interactions where the multiple-spin interactions are supposed to be weak. 
The concept of such a frustration in the nature of itinerant magnets is referred to as itinerant frustration~\cite{Hayami_PhysRevB.103.054422,hayami2021topological}.

In the present study, we propose yet another interesting situation that arises from itinerant frustration where $\chi_{\bm{q}}$ shows distinct multiple peaks at several wave vectors that are {\it not} symmetry-related with each other. 
Specifically, we focus on the situation caused by the competing interactions at different wave vectors. 
For example, in the square-lattice case in Fig.~\ref{fig:ponti}, one can consider the situation where $\chi_{\bm{q}}$ at $\bm{q}=\bm{Q}_1=(Q,0)$ and $\bm{q}=\bm{Q}_3=(Q',Q')$ take similar values, $\chi_{\bm{Q}_1}\simeq \chi_{\bm{Q}_3}$, but they are not connected by the symmetry operation.  
We find that such a competition gives rise to multiple SkXs with the skyrmion number $n_{\rm sk}$ of one and two based on simulated annealing for an effective spin model of the Kondo lattice model on a square lattice. 
We show that the competing interactions at different $\bm{q}$ lead to the rectangle SkX with $n_{\rm sk }=1$ and the square SkX with $n_{\rm sk}=2$ in addition to the square SkX with $n_{\rm sk }=1$. 
We also obtain various multiple-$Q$ states distinct from the SkXs depending on the strength of the multiple-spin interaction and magnetic field. 
Our result will stimulate further exploration of multiple-skyrmion-hosting materials in itinerant magnets.

We consider an effective spin model of the Kondo lattice model on the square lattice, which is given by 
\begin{align}
\label{eq:Model}
\mathcal{H}=  2\sum_\nu
\left( -J  \lambda_\nu
+\frac{K}{N} 
\lambda_\nu^2 
\right)-H \sum_i S_i^z,
\end{align}
where $\lambda_\nu=\sum_{\alpha,\beta}\Gamma^{\alpha\beta}_{\bm{Q}_{\nu}} S^\alpha_{\bm{Q}_{\nu}} S^\beta_{-\bm{Q_{\nu}}} $ for $\alpha,\beta=x,y,z$, $\bm{S}_{\bm{Q}_\nu}$ is the Fourier transform of the classical localized spin $\bm{S}_i$, and $N$ is the system size. 
The first term represents the bilinear (RKKY) interaction $J$ and the second term represents the biquadratic interaction $K$ defined in momentum space. 
The interactions are derived from the perturbative expansion of the Kondo lattice model in terms of $J_{\rm K}$; $J \propto J^2_{\rm K}$ and $K \propto J^4_{\rm K}$~\cite{Hayami_PhysRevB.95.224424}. 
We here neglect the other four-spin interactions between the different ordering vectors for simplicity. 
It is noted that the biquadratic interaction also arises from the order-by-disorder effect arising from thermal and/or quantum fluctuations~\cite{Okubo_PhysRevLett.108.017206,Kamiya_PhysRevX.4.011023} and short-ranged multi-spin interactions~\cite{takahashi1977half,yoshimori1978fourth,Momoi_PhysRevLett.79.2081,heinze2011spontaneous,Hoffmann_PhysRevB.101.024418}. 
These effects can contribute to $K$ at a level of the effective Hamiltonian in Eq.~(\ref{eq:Model}) when regarding the model parameters as phenomenological ones.

The wave vector $\bm{Q}_\nu$ are set by the nesting of the Fermi surfaces. 
We suppose the situation where the bare susceptibility $\chi_{\bm{q}}$ shows the first maxima at $\bm{Q}_1=(2\pi/5,0)$ and $\bm{Q}_2=(0, 2\pi/5)$ and the relatively large values at $\bm{Q}_3=(\pi/5,\pi/5)$ and $\bm{Q}_4=(-\pi/5, \pi/5)$ satisfying $\bm{Q}_1=\bm{Q}_3-\bm{Q}_4$ and $\bm{Q}_2=\bm{Q}_3+\bm{Q}_4$, as schematically shown in Fig.~\ref{fig:ponti}(a). 
The wave vectors $\bm{Q}_1$ and $\bm{Q}_2$ are higher harmonics of $\bm{Q}_3$ and $\bm{Q}_4$, which is a source of multiple SkXs as described below.
We note that the interactions at the other wave vectors $\bm{q}'$ are not important to determine the ground-state spin configuration under the condition $\chi_{\bm{q}'} < \chi_{\bm{Q}_1}, \chi_{\bm{Q}_2}, \chi_{\bm{Q}_3}, \chi_{\bm{Q}_4}$.
In real materials, the Fermi surfaces connected by $\bm{Q}_1$-$\bm{Q}_4$, which give rise to a distinct peak structure of $\chi_{\bm{Q}_1}$-$\chi_{\bm{Q}_4}$, would be promising to realize the present situation. 

For these ordering vectors, the interaction tensors $\Gamma^{\alpha\beta}_{\bm{Q}_\nu}$ to satisfy the tetragonal lattice symmetry are given by 
$\Gamma^{yy}_{\bm{Q}_1}=\Gamma^{xx}_{\bm{Q}_2}=\gamma_1$, 
$\Gamma^{xx}_{\bm{Q}_1}=\Gamma^{yy}_{\bm{Q}_2}=\gamma_2$, 
$\Gamma^{zz}_{\bm{Q}_1}=\Gamma^{zz}_{\bm{Q}_2}=\gamma_3$, 
$\Gamma^{xx}_{\bm{Q}_3}=\Gamma^{yy}_{\bm{Q}_3}=\Gamma^{xx}_{\bm{Q}_4}=\Gamma^{yy}_{\bm{Q}_4}=\gamma_4$, 
$-\Gamma^{xy}_{\bm{Q}_3}=-\Gamma^{yx}_{\bm{Q}_3}=\Gamma^{xy}_{\bm{Q}_4}=\Gamma^{yx}_{\bm{Q}_4}=\gamma_5$, 
$\Gamma^{zz}_{\bm{Q}_3}=\Gamma^{zz}_{\bm{Q}_4}=\gamma_6$ (the others are zero), which is obtained by the perturbative expansion from the Kondo lattice model~\cite{Hayami_PhysRevLett.121.137202,yambe2021skyrmion}.
Although the magnitude and sign of the anisotropic form factors are determined by the spin-orbit coupling, the basis wave function, and the Fermi surface geometry, we choose them phenomenologically as follows.
We set $J=1$ as the energy unit of the model, and choose the anisotropic parameters $\gamma_1=0.9$, $\gamma_2=0.855$, $\gamma_3=1$, $\gamma_4=0.81$, $\gamma_5=0.06525$, and $\gamma_6=0.9$, where $\gamma_1$ and $\gamma_2$ ($\gamma_4$ and $\gamma_5$) stand for the in-plane bond-dependent anisotropy, while $\gamma_3$ ($\gamma_6$) denotes the easy-axis anisotropy at $\bm{Q}_1$ and $\bm{Q}_2$ ($\bm{Q}_3$ and $\bm{Q}_4$).  
For $\gamma_1=\gamma_2=\gamma_3$, $\gamma_4=\gamma_6$, and $\gamma_5=0$, the model reduces to the isotropic spin model. 
We set the anisotropic parameters so as to satisfy $\chi_{\bm{Q}_1}=\chi_{\bm{Q}_2} >\chi_{\bm{Q}_3}=\chi_{\bm{Q}_4}$, which means that the helix with $\bm{Q}_1$ or $\bm{Q}_2$ has a smaller energy than that with $\bm{Q}_3$ or $\bm{Q}_4$.
The last term in Eq.~(\ref{eq:Model}) represents the Zeeman coupling to an external magnetic field $H$. 

\begin{figure}[t!]
\begin{center}
\includegraphics[width=1.0\hsize]{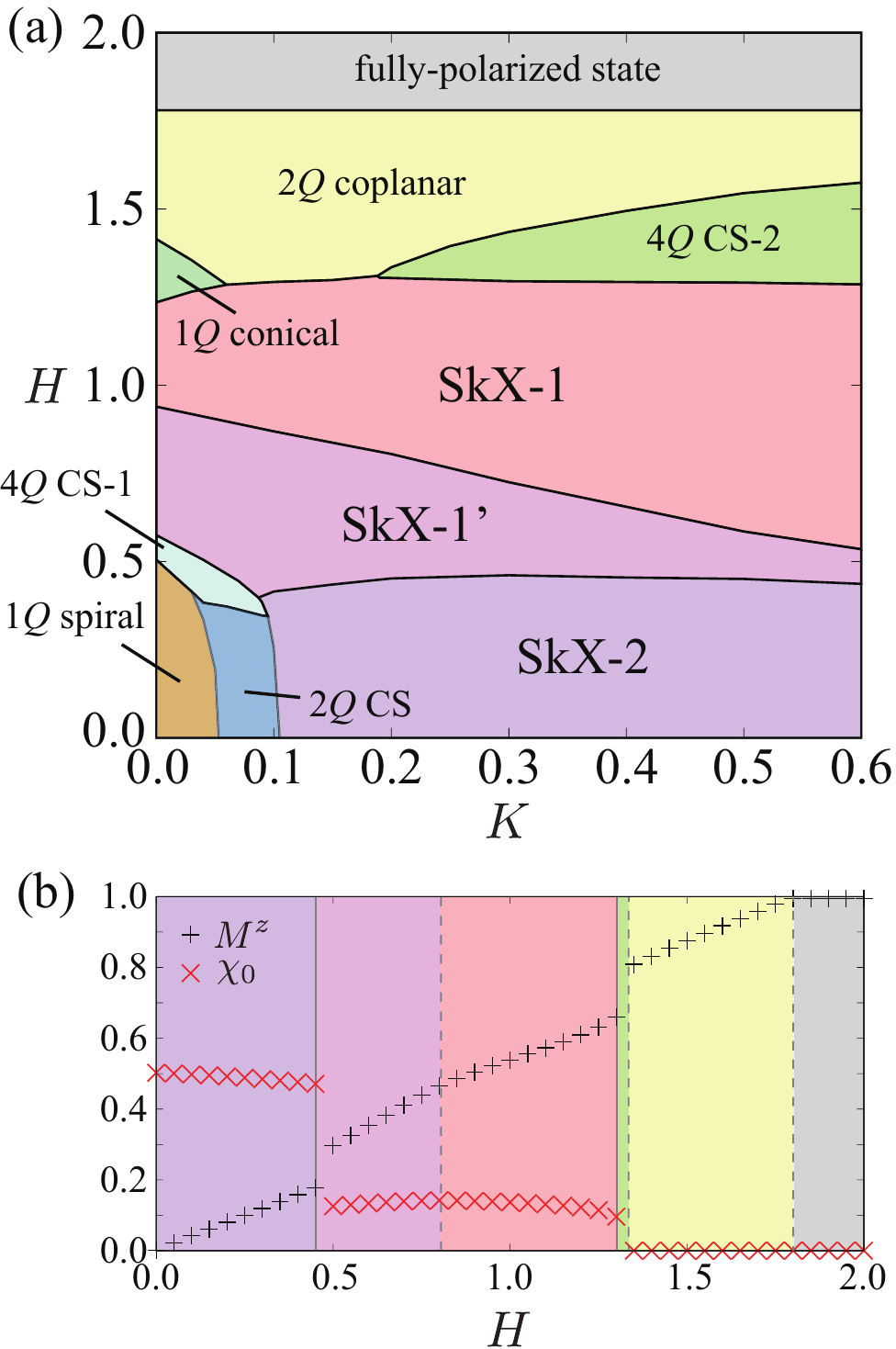} 
\caption{
\label{fig:PD}
(a) Phase diagram of the model in Eq.~(\ref{eq:Model}) obtained by simulated annealing. 
SkX and CS represent the skyrmion crystal and chiral stripe, respectively. 
(b) $H$ dependence of $M^z$ and $\chi_0$ at $K=0.2$. 
The vertical solid and dashed lines represent the topological and non-topological transitions, respectively. 
}
\end{center}
\end{figure}

It was shown that the parameter set of $(\gamma_1, \gamma_2, \gamma_3)$ for $\gamma_4=\gamma_5=\gamma_6=0$ is enough to stabilize the square SkX with $n_{\rm sk}=1$ under $H$ [denoted as SkX-1~\cite{Hayami_PhysRevB.103.024439} in Fig.~\ref{fig:PD}(a)], where the essence is lied in choosing the easy-axis anisotropy $\gamma_3> \gamma_1, \gamma_2$, which tends to stabilize the square SkX. 
The difference between $\gamma_1$ and $\gamma_2$ is introduced so as to fix the spiral plane, which can be taken to be negligibly small. 
This parameter set of $(\gamma_1, \gamma_2, \gamma_3)$ well reproduces the experimental observations of the SkX and the other multiple-$Q$ states in a skyrmion-hosting material GdRu$_2$Si$_2$~\cite{Yasui2020imaging}. 
The other model parameters $(\gamma_4,\gamma_6)$ are chosen to realize the situation with the competing interactions in momentum space as $\gamma_4/\gamma_1=\gamma_6/\gamma_3=0.9$ satisfying $\chi_{\bm{Q}_1} >\chi_{\bm{Q}_3}$. 
The following results are, at least, qualitatively similar to $0.8 \lesssim \gamma_4/\gamma_1, \gamma_6/\gamma_3<1$. 
The remaining parameter $2\gamma_5/(1-\gamma_2)=0.9$ is chosen to fix the spiral plane.

We study the magnetic phase diagram of the model in Eq.~(\ref{eq:Model}) by simulated annealing~\cite{Hayami_PhysRevB.95.224424,hayami2020multiple}. 
Our simulations are carried out with the Metropolis local updates for $\bm{S}_i$ in real space~\cite{metropolis1953equation}.
In each simulation, starting from a random spin configuration at a high temperature, $T_0 = 1$-$10$, we gradually reduce the temperature at the rate of $T_{n+1} = \alpha T_n$ to obtain the lowest-energy state, where $T_n$ is the temperature in the $n$th step and $\alpha = 0.99995$-$0.99999$. 
The final temperature is typically taken as $T = 0.01$, which is reached by a total of $10^5$-$10^6$ Monte Carlo sweeps. 
At the target temperature, we perform $10^5$-$10^6$ Monte Carlo sweeps after equlibration. 
We also start the simulations from the spin configurations obtained at low temperatures to determine the phase boundaries.
In the following, we present the results for $N=100^2$.

\begin{figure}[t!]
\begin{center}
\includegraphics[width=0.95\hsize]{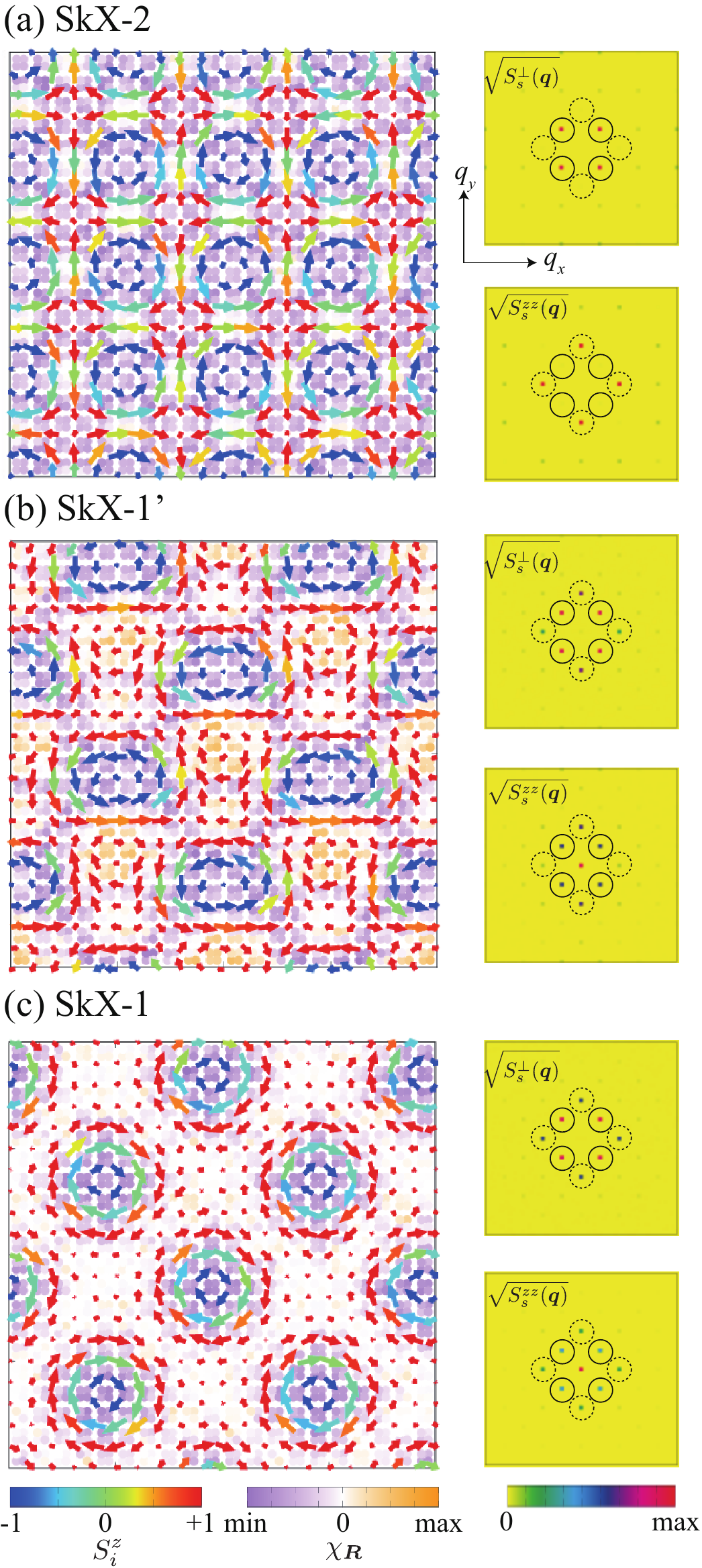} 
\caption{
\label{fig:chiral}
Snapshots of the spin configurations in (a) the SkX-2 for $K=0.2$ and $H=0.2$, (b) the SkX-1' for $K=0.2$ and $H=0.6$, and (c) the SkX-1 for $K=0.2$ and $H=1$. 
The arrows and their color show $(S^x_i, S^y_{i})$ and $S^z_i$, respectively. 
The scalar chirality $\chi_{\bm{R}}$ is also shown. 
The $xy$ and $z$ components of the spin structure factor, $\sqrt{S^{\perp}_s (\bm{q})}$ and $\sqrt{S^{zz}_s (\bm{q})}$, in the first Brillouin zone are shown in the right panel. 
The open (dashed) circles represent the positions at $\bm{Q}_3=(\pi/5, \pi/5)$ and $\bm{Q}_4=(-\pi/5, \pi/5)$ [$\bm{Q}_1=(2\pi/5, 0)$ and $\bm{Q}_2=(0, 2\pi/5)$]. 
}
\end{center}
\end{figure}

Figure~\ref{fig:PD}(a) shows the phase diagram while varying $K$ and $H$. 
Notably, we find three SkX phases in the phase diagram. 
The first SkX appears in the region for $K \gtrsim 0.1$ and $0 \leq H \lesssim 0.45$, which is denoted as the SkX-2. 
In this state, the spin configuration is characterized by the double-$Q$ peaks with equal intensity in the $xy$ component of the spin structure factor $S^{\perp}_s(\bm{q})=(1/N)\sum_{i,j}(S^x_iS^x_j+S^y_iS^y_j)e^{i\bm{q}\cdot (\bm{r}_i-\bm{r}_j)}$ at $\bm{Q}_3$ and $\bm{Q}_4$ and the $z$ component of the spin structure factor $S^{zz}_s(\bm{q})=(1/N)\sum_{i,j}S^z_iS^z_j e^{i\bm{q}\cdot (\bm{r}_i-\bm{r}_j)}$ at $\bm{Q}_1$ and $\bm{Q}_2$, as shown in Fig.~\ref{fig:chiral}(a). 
This spin configuration is regarded as the superposition of four sinusoidal waves as $\bm{S}_i=(1/N_i) [a_{xy} (-\sin \mathcal{Q}_3 +\sin \mathcal{Q}_4),a_{xy}( \sin \mathcal{Q}_3 +\sin \mathcal{Q}_4),a_z (\cos \mathcal{Q}_1 +\cos \mathcal{Q}_2)]$, where $\mathcal{Q}_\nu=\bm{Q}_\nu \cdot \bm{r}_i+\theta_\nu$ ($\theta_\nu$ is the phase of waves), $a_{xy}$ and $a_z$ are the numerical coefficients, and $N_i$ is the normalization constant.  
In the real-space picture, there are two pairs of merons with the opposite vorticity but the same scalar chirality in a magnetic unit cell, which results in $n_{\rm sk}=\pm 2$. 
This is why we call this state the SkX-2. 
The degeneracy of $n_{\rm sk}$ is owing the symmetry of the transformation of $\bm{Q}_3 \to - \bm{Q}_3$ or $\bm{Q}_4 \to - \bm{Q}_4$ in the spin configuration. 
We show the contour of the scalar chirality $\chi_{\bm{R}}=\bm{S}_i \cdot (\bm{S}_j \times \bm{S}_k)$ in Fig.~\ref{fig:chiral}(a), whose summation in the magnetic unit cell is related to $n_{\rm sk}$. 
Although a similar magnetic texture has recently been discussed in frustrated magnets with the competing interactions in momentum space~\cite{Wang_PhysRevB.103.104408}, the present SkX-2 is the first observation based on itinerant tetragonal magnets.

The second SkX phase is stabilized in the intermediate-$H$ region, next to the SkX-2 phase upon increasing $H$. 
This state exhibits the dominant peaks with equal intensity at $\bm{Q}_3$ and $\bm{Q}_4$ in both $S^{\perp}_s(\bm{q})$ and $S^{zz}_s(\bm{q})$, as shown in Fig.~\ref{fig:chiral}(b). 
In contrast to the SkX-2, the peak intensities at $\bm{Q}_1$ and $\bm{Q}_2$ are different, which indicates the breaking of fourfold rotational symmetry. 
We call this state the SkX-1'. 
Indeed, the real-space spin configuration shows a rectangle alignment of the skyrmion core, as shown in Fig.~\ref{fig:chiral}(b). 
This state exhibits $n_{\rm sk}= -1$, where the sign of $n_{\rm sk}$ is determined by $\gamma_1$, $\gamma_2$, and $\gamma_5$~\cite{Hayami_doi:10.7566/JPSJ.89.103702}. 
The spin configuration is mainly constructed from the superposition of the spirals at $\bm{Q}_3$ and $\bm{Q}_4$, where the spiral planes are tilted from the plane perpendicular to $\bm{Q}_3$ and $\bm{Q}_4$ so as to have a more $y$-spin component, and the sinusoidal wave at $\bm{Q}_2$. 

The increase of $H$ in the SkX-1' phase drives the phase transition to the SkX-1 phase, as shown in Fig.~\ref{fig:PD}(a). 
Although this state exhibits $n_{\rm sk}=-1$ as well, the spin texture is characterized by the fourfold-symmetric double-$Q$ structures, as shown in Fig.~\ref{fig:chiral}(c). 
The spin configuration is well approximated by superposing the proper-screw spirals with $\bm{Q}_1$-$\bm{Q}_4$. 
The SkX-1 phase is stabilized even without the contributions at $\bm{Q}_3$ and $\bm{Q}_4$~\cite{Hayami_PhysRevB.103.024439}, which also appears in the frustrated magnets with the bond-dependent anisotropy~\cite{Utesov_PhysRevB.103.064414,Wang_PhysRevB.103.104408} and double-exchange model with the antisymmetric spin-orbit coupling~\cite{Kathyat_PhysRevB.102.075106}.

Among the three SkXs, the emergence of the SkX-2 and the SkX-1' is owing to the competing interactions at $\bm{Q}_1$-$\bm{Q}_4$. 
In particular, the latter SkX-1' is, to the best of our knowledge, the first realization that has never been reported. 
It is also remarkable that the SkX-1' and the SkX-1 are stable even without the multiple-spin interactions, i.e., $K=0$, where the model has only the bilinear interaction and reduces to the RKKY model and Heisenberg model. 
This is attributed to the energy gain in the SkX from the contribution of higher harmonics, i.e., $\bm{Q}_1=\bm{Q}_3-\bm{Q}_4$ and $\bm{Q}_2=\bm{Q}_3+\bm{Q}_4$. 
Thus, our result indicates that the competing interactions in the symmetry-unrelated ordering vectors might be important to examine the stability of the SkXs or other multiple-$Q$ states with higher harmonics. 

Another interesting feature is that the phase transitions between the three SkXs are induced by the external magnetic field, which is in contrast to those between the two SkXs in previous studies~\cite{Ozawa_PhysRevLett.118.147205,Hayami_PhysRevB.99.094420,amoroso2020spontaneous,Wang_PhysRevB.103.104408,yambe2021skyrmion}. 
While changing the magnetic field, the phase change between the SkX-2 and SkX-1' can be easily observed by jumps of the magnetization $M^z=(1/N)\sum_i S_i^z$ and the net scalar chirality $\chi_0=[ (1/N)
\sum_{i,\delta=\pm1}\bm{S}_i \cdot (\bm{S}_{i+\delta \hat{x}}\times \bm{S}_{i+\delta \hat{y}})]^2$, where $\hat{x}$ ($\hat{y}$) is the unit vector in the $x$ ($y$) direction, as shown in Fig.~\ref{fig:PD}(b). 
Meanwhile, there is no clear anomaly in the phase transition between the SkX-1' and the SkX-1 in Fig.~\ref{fig:PD}(b). 
In this case, however, one can distinguish them from the symmetry viewpoint, since the SkX-1 holds fourfold rotational symmetry, while the SkX-1' does not.

\begin{figure}[t!]
\begin{center}
\includegraphics[width=1.0\hsize]{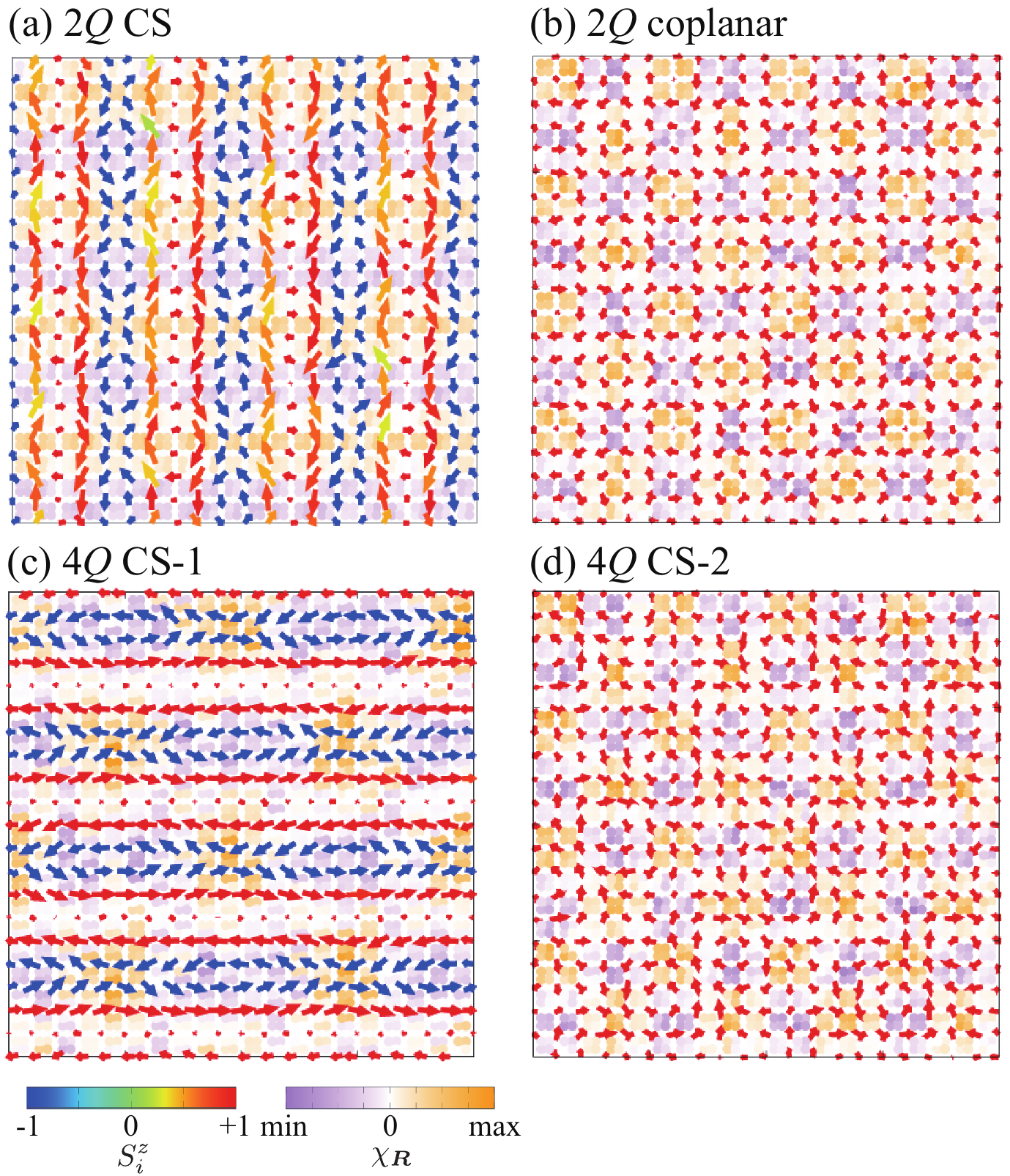} 
\caption{
\label{fig:nochiral}
Snapshots of the spin configurations in (a) the 2$Q$ CS for $K=0.1$ and $H=0$, (b) the 2$Q$ coplanar for $K=0.2$ and $H=1.5$, (c) the 4$Q$ CS-1 for $K=0$ and $H=0.55$, and (d) the 4$Q$ CS-2 for $K=0.4$ and $H=1.4$. 
The arrows and their color show $(S^x_i, S^y_{i})$ and $S^z_i$, respectively. 
The scalar chirality $\chi_{\bm{R}}$ is also shown. 
}
\end{center}
\end{figure}

The competing interactions at $\bm{Q}_1$-$\bm{Q}_4$ also give rise to unconventional multiple-$Q$ states with the chirality density waves but without the skyrmion number, $n_{\rm sk}=0$, in addition to the single-$Q$ (1$Q$) conical state with $\bm{Q}_1$ or $\bm{Q}_2$ (the spiral plane is the $xy$ plane) and 1$Q$ spiral state with $\bm{Q}_1$ or $\bm{Q}_2$ (the spiral plane is perpendicular to $\bm{Q}_1$ and $\bm{Q}_2$). 
We find four multiple-$Q$ states: $2Q$ chiral stripe (CS), $2Q$ coplanar, 4$Q$ CS-1, and 4$Q$ CS-2 states, as shown in Fig.~\ref{fig:PD}(a). 
The 2$Q$ CS state for small $K$ and small $H$ is described by the superposition of the single-$Q$ spiral along the $\bm{Q}_1$ direction and the single-$Q$ sinusoidal wave along the $\bm{Q}_2$ direction~\cite{Ozawa_doi:10.7566/JPSJ.85.103703,yambe2020double}, while the 2$Q$ coplanar state for large $H$ is by that of the two sinusoidal waves along the $\bm{Q}_1$ and $\bm{Q}_2$ directions, which have been found in the itinerant electron model without the higher-harmonic contributions~\cite{Hayami_PhysRevB.103.024439} and the frustrated spin model without the multiple-spin interactions~\cite{Utesov_PhysRevB.103.064414,Wang_PhysRevB.103.104408}. 
The spin and chirality configurations of the 2$Q$ CS and 2$Q$ coplanar states are shown in Figs.~\ref{fig:nochiral}(a) and \ref{fig:nochiral}(b), respectively. 
The other 4$Q$ states are a consequence of the present model with competing interactions. 
The 4$Q$ CS-1 state for small $K$ and intermediate $H$ resembles the 2$Q$ CS state but have additional sinusoidal modulations along the $\bm{Q}_3$ and $\bm{Q}_4$ directions. 
The 4$Q$ CS-2 state for large $K$ and large $H$ is described by a superposition of three inplane spirals at $\bm{Q}_1$, $\bm{Q}_2$, and $\bm{Q}_3$ and almost inplane spiral slightly tilted to have the $z$ spin component at $\bm{Q}_4$. The spin and chirality configurations in both 4$Q$ states are presented in Figs.~\ref{fig:nochiral}(c) and \ref{fig:nochiral}(d). 

To summarize, we found that the competing interactions arising from the multiple peaks in the bare susceptibility in itinerant magnets stabilize multiple SkXs with different skyrmion numbers. 
The competing interactions at the RKKY level are enough to stabilize the SkX with $n_{\rm sk}=1$ in tetragonal itinerant magnets. 
Meanwhile, it was shown that the biquadratic interaction leads to the SkX with $n_{\rm sk}=2$ at zero field. 
Our argument based on the effective spin model in Eq.~(\ref{eq:Model}) holds for arbitrary $|\bm{Q}|$ and for any tetragonal systems. 
We also showed that a variety of multiple-$Q$ states appear by considering the competing interactions in momentum space. 
Our study will provide rich multiple-$Q$ spin textures that emerge from itinerant frustration, which are relevant with the recent experimental findings of the SkX and the other multiple-$Q$ phases in GdRu$_2$Si$_2$~\cite{khanh2020nanometric,Yasui2020imaging} and EuAl$_4$~\cite{Shang_PhysRevB.103.L020405,kaneko2021charge}. 
Moreover, as the present scenario based on itinerant frustration can happen in the other topological spin textures in the other lattice structures, such as the meron-antimeron crystal on the triangular lattice~\cite{Hayami_PhysRevB.104.094425} and the hedgehog crystal on the cubic lattice~\cite{Okumura_PhysRevB.101.144416,Kato_PhysRevB.104.224405}, it is interesting to explore further intriguing topological spin textures and their related phase transitions.

\begin{acknowledgments}
This research was supported by JSPS KAKENHI Grants Numbers JP19K03752, JP19H01834, JP21H01037, and by JST PRESTO (JPMJPR20L8). 
Parts of the numerical calculations were performed in the supercomputing systems in ISSP, the University of Tokyo.
\end{acknowledgments}

\bibliographystyle{JPSJ}
\bibliography{ref}

\end{document}